\documentstyle[12pt,epsfig,subfigure,amsfonts,amssymb]{article}

\newcommand{\be}{\begin{equation}}
\newcommand{\ee}{\end{equation}}
\newcommand{\beq}{\begin{equation}}
\newcommand{\eeq}{\end{equation}}
\newcommand{\ba}{\begin{eqnarray}}
\newcommand{\ea}{\end{eqnarray}}
\newcommand{\bea}{\begin{eqnarray}}
\newcommand{\eea}{\end{eqnarray}}

\begin{document}
\baselineskip=15.5pt \pagestyle{plain} \setcounter{page}{1}
%--------+---------+---------+---------+---------+---------+---------+
%--------+---------+---------+---------+---------+---------+---------+
\begin{titlepage}

\vskip 0.8cm

\begin{center}

{\LARGE Quantum corrections to dynamical holographic thermalization:
entanglement entropy and other non-local observables} \vskip .3cm

\vskip 1.cm

{\large {Walter H. Baron{\footnote{\tt wbaron@fisica.unlp.edu.ar}} and
Martin Schvellinger{\footnote{\tt martin@fisica.unlp.edu.ar}}}}

\vskip 0.5cm

{\it IFLP-CCT-La Plata, CONICET and Departamento  de F\'{\i}sica,
Universidad Nacional de La Plata.  Calle 49 y 115, C.C. 67, (1900)
La Plata,  Buenos Aires,
Argentina.} \\

\vspace{1.cm}

{\bf Abstract}

\end{center}

We investigate the thermalization time scale in the planar limit of
the $SU(N)$ ${\cal {N}}=4$ SYM plasma at strong yet finite 't Hooft
coupling by considering its supergravity dual description, including
the full ${\cal {O}}(\alpha'^3)$ type IIB string theory corrections.
We also discuss on the effects of the leading non-planar
corrections. We use extended geometric probes in the bulk which are
dual to different non-local observables in the ${\cal {N}}=4$ SYM
theory. This is carried out within the framework of dynamical
holographic thermalization.

\noindent

\end{titlepage}

\newpage

\tableofcontents

\vfill

\newpage

%%%%%%%%%%%%%%%%%%%%%%%%%%%%%%%%%%%%%%%%%%%%%%%%%%%%%%%%%%%%%%%%%%%%%%%
\section{Introduction and motivation}
%%%%%%%%%%%%%%%%%%%%%%%%%%%%%%%%%%%%%%%%%%%%%%%%%%%%%%%%%%%%%%%%%%%%%%%

Heavy-ion collision experiments at the Relativistic Heavy Ion
Collider (RHIC) and the Large Hadron Collider (LHC) lead to the
formation of a state of QCD matter known as quark-gluon plasma
(QGP). It has been shown that the hydrodynamical behavior of such a
system is compatible with low values of the viscosity. This is an
indication that the referred QGP is strongly coupled. Therefore,
perturbative quantum field theory methods are not suitable to
investigate this plasma. This is where the gauge/string duality
enters, since it allows one to describe properties of a strongly
coupled gauge theory in terms of its dual gravitational description.
In particular, a holographic dual pair which has been very useful to
understand important results of heavy-ion collider physics is given
by the planar limit of the $SU(N)$ ${\cal {N}}=4$ supersymmetric
Yang-Mills theory (SYM), and by type IIB supergravity theory on the
anti-de Sitter Schwarzschild black hole times a $S^5$ background.
Motivated by QCD and $SU(N)$ ${\cal {N}}=4$ SYM lattice
calculations, it has been argued (see for instance
\cite{CasalderreySolana:2011us} and references therein) that for QGP
equilibrium temperatures which are just above the QCD deconfinement
temperature, several properties of both gauge theories behave in a
similar way, even quantitatively. This supports the idea of using
$SU(N)$ ${\cal {N}}=4$ SYM theory, which is much more symmetric and
therefore a much easier theory to work with than the strongly
coupled regime of QCD, in terms of its dual string theory model at
finite temperature in order to describe the strongly coupled plasma.

In an actual collision of two heavy nuclei part of their kinetic
energy in the center of mass frame transforms into intense heat
before the system reaches the thermal equilibrium, leading to a
strongly coupled plasma. When the collision occurs the initial state
represents a far-from-equilibrium system. A proposal to model such a
system has been the so-called holographic thermalization, where it
is assumed that there is a sudden injection of energy to the system
which in the thermal quantum field theory is interpreted as a
thermal quench. The time evolution of the thermalizing system is a
very complicated process to model due to the large number of degrees
of freedom involved as well as its non-perturbative character.

We consider a model of dynamical holographic thermalization
\cite{AbajoArrastia:2010yt,Albash:2010mv,Ebrahim:2010ra,Balasubramanian:2010ce,
Balasubramanian:2011ur,Garfinkle:2011hm,Aparicio:2011zy,Allais:2011ys,Keranen:2011xs,
Garfinkle:2011tc,Das:2011nk,Hubeny:2012ry,Galante:2012pv,Caceres:2012em,
Wu:2012rib,Baron:2012fv} which has been recently used to study
thermalization time scale of strongly coupled plasmas. Although most
of these applications have been done by using effective holographic
dual models, in the present work we focus on an specific string
theory dual model in order to investigate quantum corrections to the
holographic thermalization time scale of a particular strongly
coupled SYM plasma whose string dual description is very well known.
We can schematically describe this thermalization model starting
from an initial vacuum gravity solution given by an anti-de Sitter
spacetime (AdS), which represents the holographic dual description
of a certain quantum field theory at zero temperature. The final
state is defined by an asymptotically anti-de Sitter Schwarzschild
black hole (AdS-BH), with Hawking temperature $T_H$, which is the
gravity dual model corresponding to the quantum field theory at
finite equilibrium temperature $T$. Furthermore, it is assumed that
$T=T_H$. Interestingly, there is a solution of the Einstein
equations which interpolates between these gravity solutions, and it
can be represented by a thin shell collapsing from the AdS-boundary.
The shell separates the space into two regions: the inner one is an
AdS and the outer one is an AdS-BH. As we shall explain, extended
probes in the bulk can be used in order to measure the
thermalization time scale of the boundary theory plasma.

The general picture which emerges from the gauge/gravity duality
when studying the dynamical evolution of a collapsing thin shell
shows that UV modes thermalize faster than IR ones. This statement
is based upon the fact that extended geometric probes in the bulk,
such as space-like geodesic curves, minimal area surfaces and
minimal volume hyper-surfaces, thermalize faster when the boundary
field theory separation between two operators (which are dual to
these bulk geometric probes) becomes shorter. Thus, the claim is
that the faster thermalization of short distance two-point functions
of quantum field theory operators of the boundary field theory is
related to faster thermalization of UV modes, in comparison with IR
modes which are associated with two-point functions of operators
with larger separation between them.

The referred behavior has been found for very different kinds of
degrees of freedom composing the shell \cite{Baron:2012fv} in
effective five-dimensional holographic dual models, and from the
gravity point of view it follows as a consequence of the
construction of the geometric probes. In particular, this also
occurs in the planar limit of the $SU(N)$ ${\cal {N}}=4$ SYM plasma
at the strong 't Hooft coupling ($\lambda$) limit. This result is
obtained from its dual supergravity description, {\it i.e.} with no
string theory corrections. Thus, a natural question which arises is
whether string theory quantum corrections modify or not that
statement about UV/IR modes thermalization, and if they do, it is
crucial to know how are such corrections. In this paper we address
this question by considering the leading string theory corrections
to type IIB supergravity. Very interestingly, we find the emergence
of an energy scale which separates the thermalization time scales of
IR and UV modes. Thus, by decreasing $\lambda$ the thermalization
time scale for IR modes slightly increases in comparison with its
thermalization time scale at the $\lambda \rightarrow \infty$ limit,
when final states are compared at fixed temperature or fixed energy,
{\it i.e.} they thermalize a bit slower than in the strong coupling
limit. On the other hand, the alluded corrections induce an opposite
behavior on the thermalization time for UV modes, {\it i.e.} they
thermalize slightly faster compared with the strong coupling limit.
We shall discuss about these two distinctive effects in the last
section of the paper. We also consider the effect of string-loop
corrections which lead to $1/N$ corrections in the dual SYM theory,
and show that their effects go in the same direction as for the
mentioned higher curvature corrections.

In particular, we study the dynamical evolution of a thin shell
composed by massless type IIB supergravity degrees of freedom,
collapsing within the asymptotically AdS spaces described before.
This is aimed at investigating holographic thermalization of $SU(N)$
${\cal {N}} = 4$ SYM theory plasma at strong yet finite coupling. On
the boundary quantum field theory this corresponds to considering a
certain thermal quench. As we have already commented, we assume it
corresponds to a sudden injection of energy due to a heavy-ion
collision. Essentially, this energy is a fraction of the center of
mass kinetic energy of the two colliding heavy ions which is
transformed into intense heat during the plasma thermalization. In a
certain way this energy enters the definition of the parameter $M$
introduced in the thermal quench. We study the thermalization time
scale by calculating renormalized space-like geodesic lengths,
rectangular and circular minimal area surfaces, and three-dimensional
minimal volume hyper-surfaces, as extended probes of thermalization,
which are claimed to be related to two-point functions,
rectangular and circular Wilson loops, and entanglement entropy,
respectively. Thus, on the quantum field theory side we consider
different non-local observables to probe thermalization of this
strongly coupled system.

We consider three scenarios described in terms of the parameters of
the model, which are the equilibrium temperature $T$ and the energy
parameter $M$ of the quench, which are related each other by
Eq.(\ref{temp}). One situation is when we fix the horizon radius
$z_h$ or equivalently the parameter $M$ (see Eq.(\ref{f(z)})),
independently of the quantum corrections, this makes the equilibrium
temperature to be dependent of the 't Hooft coupling. Secondly, we
consider the case when the equilibrium temperature is kept fixed,
while the horizon radius becomes a function of the 't Hooft
coupling. These two situations show the same behavior for the
thermalization, as explained before. Then, we discuss what happens
when we vary the parameter $M$ of the thermal quench in an arbitrary
way, which we associate with the fraction of the center of mass
kinetic energy of the two colliding heavy ions which transforms into
heat and initiates the thermalization process. We discuss in detail
these three cases in section 4.

Notice that recently higher curvature corrections to the $SU(N)$
${\cal {N}}=4$ SYM plasma thermalization using a quasi-static
approximation have been discussed in
\cite{Steineder:2012si,Steineder:2013ana}, obtaining a different
behavior for the UV and IR modes compared with our results. However,
one has to be cautious making that comparison since in these
references the $SU(N)$ ${\cal {N}}=4$ SYM theory is coupled to an
external electromagnetic field, which induces large ${\cal
{O}}(\alpha'^3)$ corrections for certain observables. In section 5
we shall discuss more about this in comparison with our findings.

The paper is organized as follows. In Section 2 we describe the
leading string theory corrections to the type IIB
supergravity action, including string-loop corrections. In section 3
we consider the quantum corrections to the holographic
thermalization process. Results and discussions are introduced in
section 4. Section 5 is devoted to the conclusions.

%%%%%%%%%%%%%%%%%%%%%%%%%%%%%%%%%%%%%%%%%%%%%%%%%%%%%%%%%%%%%%%%%%%%%%%%%%%%%%
\section{Leading type IIB string theory corrections}
%%%%%%%%%%%%%%%%%%%%%%%%%%%%%%%%%%%%%%%%%%%%%%%%%%%%%%%%%%%%%%%%%%%%%%%%%%%%%%

The ten-dimensional AdS$_5$-BH$\times S^5$ metric is given by
\be
ds^2 = \frac{R^2}{z^2} \left[ -f(z) \,
 dt^2 + d\vec{x}^2 + \frac1{f(z)} \,
 dz^2\right] + R^2 \, d\Omega_5^2 \, . \label{metric}
\ee
This is in fact an exact solution of type IIB supergravity, which
turns out to be the metric of the background of the holographic dual
model corresponding to the planar limit of the $SU(N)$ ${\cal
{N}}=4$ SYM theory at finite temperature, $T$, in the strong
coupling limit. In the above expression we define
\be
f(z)=1- \frac{z^4}{z_h^4}=1- 2M z^4 \, ,\label{f(z)}
\ee
while the radius of the AdS$_5$ and the five-sphere is $R$. At
$z=0$ we have the boundary of the AdS$_5$ space, while the black
hole horizon is set at $z=z_h$.

Now, let us consider the leading type IIB string theory corrections
to the supergravity action $S_{IIB}^{SUGRA}$. The corrections are
described by the term $S_{{\cal {R}}^4}^{3}$. Therefore, the total
action at ${\cal {O}}(\alpha'^3)$ can be written as follows
\be
S_{IIB}=S_{IIB}^{SUGRA} + S_{{\cal {R}}^4}^{3} \, .
\ee
Since the square root of the 't Hooft coupling is related to the
inverse of $\alpha'$, at the strong 't Hooft coupling limit ($N \gg
\lambda \gg 1$) the holographic dual model is derived from type IIB
supergravity, {\it i.e.} for $\alpha' \rightarrow 0$. In particular,
$S_{IIB}^{SUGRA}$ is composed by the Einstein-Hilbert action coupled
to the dilaton and the Ramond-Ramond five-form field strength $F_5$.
In the Einstein frame it reads
\be
S_{IIB}^{SUGRA}=\frac{1}{2 \kappa_{10}^2}\int \, d^{10}x\,
\sqrt{-G}\left[R_{10}-\frac{1}{2}\left(\partial\phi
\right)^2-\frac{1}{4.5!}\left(F_5\right)^2 \right] \, .
\label{action-10D}
\ee
By considering D3-branes in type IIB string theory, in reference
\cite{Green:2003an} the contributions from higher curvature terms at
${\cal {O}}(\alpha'^3)$, as well as perturbative $1/N$ corrections
and instanton corrections were computed. The idea is to look at a
supersymmetric completion of the $C^4$ term, where $C$ is the
ten-dimensional Weyl tensor. Thus
\be
S_{{\cal {R}}^4}^{3} = \frac{\alpha'^3 g_s^{3/2}}{32 \pi G} \int
d^{10}x \int d^{16}\theta \sqrt{-g} \, f^{(0,0)}(\tau, \bar{\tau})
[(\theta \Gamma^{mnp} \theta)(\theta \Gamma^{qrs} \theta){\cal
{R}}_{mnpqrs}]^4 + c.c. \, , \label{green-stahn}
\ee
where $\tau$ is the complex scalar field written as $\tau_1 + i
\tau_2 \equiv a + i e^{-\phi}$, where $a$ is the axion, $\phi$ is
the dilaton and $e^\phi=g_s$ is the string coupling. Recall that in
terms of the gauge/string duality $g_s \equiv 1/N$. In the above
expression $f^{(0,0)}(\tau, \bar{\tau})$ is the modular form. Using
the Weyl tensor one can define the tensor ${\cal {R}}$
\cite{Green:2003an,deHaro:2002vk,Paulos:2008tn,Myers:2008yi}
\be
{\cal {R}}_{mnpqrs} = \frac{1}{8} g_{ps} C_{mnqr} + \frac{i}{48} D_m
F^+_{npqrs} + \frac{1}{384} F^+_{mnpkl} F_{qrs}^{+\,\,\, kl} \, ,
\ee
where
\be
F^+=(1+*)F_5/2 \, .
\ee
Notice that the action (\ref{green-stahn}) has been obtained using
the fact that the physical field content of type IIB supergravity
can be arranged in a scalar superfield $\Phi(x, \theta)$, where
$\theta_a$, with $a = 1, \cdot \cdot \cdot, 16$, is a complex Weyl
spinor of $SO(1, 9)$. The matrices $\Gamma$ have been defined as
usual \cite{Paulos:2008tn}. The modular form is given by the
following expression \cite{Green:1997tv}
\be
f^{(0,0)}(\tau, \bar{\tau}) = 2 \zeta(3) \tau_2^{3/2} + \frac{2
\pi^2}{3} \tau_2^{-1/2} + 8 \pi \tau_2^{1/2} \sum_{m \neq 0, n \geq
0} \frac{|m|}{|n|} e^{2 \pi i |m n| \tau_1} K_1(2 \pi |m n| \tau_2)
\, ,
\ee
where $K_1$ is the modified Bessel function of second kind which
comes from the non-perturba\-ti\-ve D-instantons contributions. The
zeta function $\zeta(3)$ is the coefficient of the first
perturbative correction in the Eisenstein series of the modular
form. Also note that in the background we consider with $N$
coincident parallel D3 branes there are some simplifications: the
axion vanishes, thus $\tau_1 = 0$, while $\tau_2 = g_s^{-1}$. Thus,
when the string coupling is small the modular form reduces to
\be
f^{(0,0)}(\tau, \bar{\tau}) = 2 (4 \pi N)^{3/2} \left(
\frac{\zeta(3)}{\lambda^{3/2}} + \frac{\lambda^{1/2}}{48 N^2} +
\frac{e^{-8 \pi^2 N/\lambda}}{2 \pi^{1/2} N^{3/2}} \right) \, .\label{ModForm}
\ee
It has been also shown that the D3-brane solution in supergravity is
not renormalized by higher derivative terms \cite{Green:2003an}.
Previously, Banks and Green had shown that $AdS_5 \times S^5$ is a
solution to all orders in $\alpha'$ \cite{Banks:1998nr}. This is not
the case for the AdS-BH we consider, whose metric does receive
corrections at order $\alpha'^3$ as we shall see shortly.

At this point, let us also consider the large $N$ limit of the dual
$SU(N)$ ${\cal {N}}=4$ SYM theory. The finite leading 't Hooft
coupling corrections in its string theory dual description are
accounted for by the following action \cite{Myers:2008yi}
\be\label{10DWeyl}
S_{IIB}^{\alpha'}=\frac{R^6}{2 \kappa_{10}^2}\int \, d^{10}x\,
\sqrt{-G}\left[ \, \gamma e^{-\frac{3}{2}\phi} \left(C^4 +
C^3{\mathcal{T}}+C^2{\mathcal{T}}^2+C{\mathcal{T}}^3+{\mathcal{T}}^4
\right)\right] \, , \label{10d-corrected-action}
\ee
which was obtained from the action (\ref{green-stahn}) in the planar
limit of the $SU(N)$ ${\cal {N}}=4$ SYM theory. Notice that we have
written this $\alpha'^3$-corrected action
(\ref{10d-corrected-action}) in the Einstein frame. Here $\gamma
\equiv \frac{1}{8} \, \zeta(3) \, (\alpha'/R^2)^{3}$, where $R^4 = 4
\pi g_s N \alpha'^2$. Since $\lambda = g_{YM}^2 N \equiv 4 \pi g_s
N$, therefore $\gamma = \frac{1}{8} \, \zeta(3) \,
\frac{1}{\lambda^{3/2}}$. This action was computed in
\cite{Paulos:2008tn}, using the methods of \cite{Green:2005qr}.

The $C^4$ term is a dimension-eight operator, given by
\be
C^4=C^{hmnk} \, C_{pmnq} \, C_h^{\,\,\,rsp} \, C^{q}_{\,\,\,rsk} +
\frac{1}{2} \, C^{hkmn} \, C_{pqmn} \, C_h^{\,\,\, rsp} \,
C^q_{\,\,\, rsk} \, ,
\ee
where $C^{q}_{\,\,\, rsk}$ is the Weyl tensor. The tensor ${\cal
{T}}$ is much more involved and it is defined by
\begin{equation}
{\cal {T}}_{abcdef}= i\nabla_a
F^{+}_{bcdef}+\frac{1}{16}\left(F^{+}_{abcmn}F^{+}_{def}{}^{mn}-3
F^{+}_{abfmn}F^{+}_{dec}{}^{mn}\right) \, , \label{T-tensor}
\end{equation}
where the indices $[a,b,c]$ and $[d,e,f]$ are antisymmetrized in
each squared brackets, and symmetrized with respect to interchange
of $abc \leftrightarrow def$ \cite{Paulos:2008tn}.

Importantly, at finite temperature the metric only has corrections
coming from the $C^4$ term. This is so because the tensor
${\mathcal{T}}$ vanishes on the supergravity solution with no string
theory corrections \cite{Myers:2008yi}. The solution to the Einstein
equations derived from the supergravity action (\ref{action-10D}) is
an AdS$_5$-BH$\times S^5$. There are $N$ units of flux of $F_5$
through the sphere. Recall that $N$ is the rank of the gauge group
in the field theory, and on the other hand, it corresponds to the
number of parallel D3-branes whose back-reaction deforms the
space-time leading to the above metric in the near horizon limit.

In conclusion, the only part of the ${\cal {O}}(\alpha'^3)$-action
which affects the metric is the $C^4$ term. This induces the
following corrected metric obtained by
\cite{Gubser:1998nz,Pawelczyk:1998pb,deHaro:2003zd}
\be
ds^2 = \frac{R^2}{z^2} \left[-f(z) \, K^2(z) \, dt^2 + d\vec{x}^2 +
\, f^{-1}(z)P^2(z) \, dz^2 \right] + R^2 L^2(z) \, d\Omega_5^2 \,
,\label{proper-metric}
\ee
using same notation as in Eq.(\ref{metric}). $K(z)$, $P(z)$ and
$L(z)$ are given by the following expressions:
\be
K(z) = e^{\gamma \, [a(z) + 4b(z)]} \, , \quad P(z) =
e^{\gamma \, b(z)} \, , \quad L(z) =  e^{\gamma \, c(z)} \,
,
\ee
where
\ba
a(z) &=& -\frac{1625}{8} \, \left(\frac{z}{z_h}\right)^4 - 175 \,
\left(\frac{z}{z_h}\right)^8 + \frac{10005}{16} \,
\left(\frac{z}{z_h}\right)^{12} \, , \nonumber \\
b(z) &=& \frac{325}{8} \, \left(\frac{z}{z_h}\right)^4 +
\frac{1075}{32} \, \left(\frac{z}{z_h}\right)^8
- \frac{4835}{32} \, \left(\frac{z}{z_h}\right)^{12} \, , \nonumber \\
c(z) &=& \frac{15}{32} \,
\left(1+\left(\frac{z}{z_h}\right)^4\right) \,
\left(\frac{z}{z_h}\right)^8 \, .
\ea
In addition, after the leading type IIB string theory corrections
are taken into account, the dilaton field becomes
$\phi=\phi_0+\gamma \phi_1+{\mathcal O}(\gamma^2)$, where
\bea
\phi_0&=&-\log(g_s) \, ,\cr\cr
\phi_1(z)&=&-\frac{45}{8}\left(\frac{z^4}{z_h^4}+\frac12\frac{z^8}{z_h^8}+
\frac13\frac{z^{12}}{z_h^{12}}\right) \, .\label{dilaton}
\eea
The temperature of the boundary field theory, which is assumed to be
equal to the Hawking temperature of the AdS$_5$-Schwarzschild black
hole, is now corrected as \cite{Pawelczyk:1998pb}
\bea
T=\frac{(2M)^{\frac14}}{\pi R^2}\left(1+\frac{265}{16}\gamma\right)
\label{temp} \, .
\eea
Having obtained the corrected metric, dilaton, and temperature,
equations (\ref{proper-metric}), (\ref{dilaton}) and (\ref{temp}),
we now focus on the leading quantum corrections to the holographic
thermalization process.

%%%%%%%%%%%%%%%%%%%%%%%%%%%%%%%%%%%%%%%%%%%%%%%%%%%%%%%%%%%%%%
\section{Quantum corrections to holographic thermalization}
%%%%%%%%%%%%%%%%%%%%%%%%%%%%%%%%%%%%%%%%%%%%%%%%%%%%%%%%%%%%%%

\subsection{A collapsing thin shell in Anti de Sitter space}

The AdS-Vaidya solution has been widely employed to account for the
evolution towards the final equilibrium state of strongly coupled
SYM plasmas
\cite{AbajoArrastia:2010yt,Albash:2010mv,Ebrahim:2010ra,Balasubramanian:2010ce,
Balasubramanian:2011ur,Garfinkle:2011hm,Aparicio:2011zy,Allais:2011ys,Keranen:2011xs,
Garfinkle:2011tc,Das:2011nk,Hubeny:2012ry,Galante:2012pv,Caceres:2012em,
Wu:2012rib,Baron:2012fv}. This solution describes the geometry of a
pressureless thin shell composed by massless particles collapsing in
an AdS spacetime\footnote{It is worth mentioning that AdS-Vaidya
metric (\ref{Vaidyametric}) is not an exact solution of type IIB
supergravity on $AdS_5\times S^5$. Nevertheless, it was shown
\cite{Bhattacharyya:2009} that this solution appears as a good
approximation of certain field configurations.}, leading to the
formation of a black hole with negative cosmological constant at
asymptotic times.

Since the shell is composed by coherent massless degrees of freedom,
it moves at the speed of light. Thus, it is set at a constant
position in the Eddington-Filkenstein-like coordinate $v_0$, defined by
\bea
dv_0=dt-f^{-1}(z)dz \, .
\eea
The AdS part of the metric describing this geometry, in the
thin-shell limit, takes the following form
\bea
ds^2=\left\{\begin{array}{lcr}\frac{R^2}{z^2}\left[- dv_0^2-2dv_0 dz
+ d\vec{x}^{\,2}\,\right] \, ,
&&  v_0<0,  \cr\cr \frac{R^2}{z^2}\left[-f(z) dv_0^2-2dv_0 dz +
d\vec{x}^{\,2}\, \right], &&  v_0>0 \, .
            \end{array}\right.\label{Vaidyametric}
\eea
Once the corrections to the pure type IIB supergravity solution are
considered, the null dust and zero width hypotheses lead to a
geometry with the metric given by
\bea%
ds^2=\left\{\begin{array}{lcr} \frac{R^2}{z^2}\left[- dv^2-2dv dz +
d\vec{x}^{\,2}\,\right] \, , && v<0 \, ,  \cr\cr
\frac{R^2}{z^2}\left[-f(z) K^{2}(z) dv^2-2 P(z) K(z) dv dz +
d\vec{x}^{\,2}\,\right]\, ,  &&  v>0 \, ,
            \end{array}\right.
\eea
where $v$ is now defined by
\bea
dv=dt-\frac{e^{-\gamma \left[a(z)+3 b(z)\right]}}{f(z)} dz \,
.\label{defv}
\eea

\subsection{Probing thermalization: non-local observables}

In order to probe the thermalization one must consider the evolution
of different non-local observables. In the AdS/CFT correspondence
such boundary quantum field theory observables are typically
identified with geometric objects in the bulk, {\it e.g.} the
Wightman two-point function of scalar field operators is determined
by the sum of curve lengths ending at the location of the operators.
In the limit of highest conformal weights, the saddle-point
approximation reduces the sum just to the contribution of the
shortest length, {\it i.e.} the geodesic length
\cite{Balasubramanian:1999}. Similarly, in the classical limit, the
Wilson loops are determined by minimal area surfaces ending on the
closed Wilson path \cite{Maldacena:1998im}, and the entanglement
entropy of a given volume in the boundary theory is set by the
minimal bulk hyper-surface ending in such a region
\cite{Ryu:2006,Hubney:2008}.

Equation (\ref{defv}) determines the position of the shell as a
function of the boundary time, $t$. Without lost of generality we
can set $t=0$ as the moment the shell leaves the boundary. Thus, a
boundary observer feels that for any $t>0$ there is a spatial scale
$\tilde l$ at the boundary such that the extended geometric objects
in the bulk (like geodesic curves and hyper-surfaces) with ends at
the boundary whose separation is smaller than $\tilde l$ are
completely embedded in the AdS-BH region of the bulk. Therefore, the
observable associated quantities of the boundary theory exactly
match their corresponding measures of the plasma in thermal
equilibrium.

On the other hand, extended objects probing larger scales than
$\tilde l$ on the boundary have to wait more time before they agree
with their corresponding thermally equilibrated values. This simple
geometric picture gives us a very simple (gravity) intuition of why,
contrary to what happens in perturbative theories, UV modes
thermalize before than the IR ones do.

\bigskip

Once quantum corrections are included non-local observables may no
longer be given by the geometric quantities described
above\footnote{We thank Robert Myers for some comments about this
point.}. Nevertheless, there is evidence supporting the statement
that the new definition for these non-local observables is not given
in terms of the length of the geodesic or the area/volume of the
minimal hypersurface, but in terms of a different functional over
the same geodesic or minimal surfaces. Thus, provided that we
exclusively focus on thermalization times, and {\it not} on the
precise value of the non-local observables, we still have to
consider the same kind of objects considered before the introduction
of $\alpha'$ corrections.

For instance, let us consider a scalar field coupled to gravity
through a standard interaction term
\bea
S_{Int}(g,\phi)=\frac12\int d^Dx \sqrt{-g}\left(\partial_\mu\phi
\partial_\nu\phi+m^2\phi^2\right) \, .
\eea
Then, following \cite{witten:98} with the new effective supergravity
action, the two-point function is now given by the old one with the
replacement of the AdS-Schwarzschild metric by the corrected one
(\ref{proper-metric}). This is up to an overall factor containing
the Weyl tensor (\ref{10d-corrected-action}), which plays no role in
the variation with respect to the boundary values of the scalar
fields. Thus, following \cite{Balasubramanian:1999}, the corrected
two-point function in the large conformal weight limit is, up to an
overall constant, given by the exponential of the geodesic length in
the space whose metric is defined by equation (\ref{proper-metric}).

Concerning the entanglement entropy, at the present it is not known
a general definition for it if one considers higher derivative
gravity. Some progress in this direction was achieved for
Gauss-Bonnet and Lovelock gravities. In this situation, it was shown
in \cite{MyersSmolkin:2011}, that Wald's formula used by
\cite{Ryu:2006} does not satisfy the subadditivity condition when
higher derivative terms are included in the Einstein-Hilbert action.
They found a natural generalization which replaces the projection of
the curvature tensor by the intrinsic curvature tensor of the
surface in the Wald's formula. This new definition was used for the
Gauss-Bonnet gravity, as well as in the next order in derivatives
(see (C.21) and (C.28) of \cite{MyersSmolkin:2011}) of Lovelock
gravity. These definitions reproduce the expressions derived in
\cite{MyersCasini:2011} for an arbitrary higher curvature theory,
and although they do not agree with the Ryu-Takayanagi (RT)
prescription, these new functionals are computed over the same
minimal surfaces used in the RT prescription (see (C.14) of
\cite{MyersSmolkin:2011}).

Recent studies \cite{MyersDamian:2013} have found that entanglement
entropy at $T=0$ does not receive corrections when leading higher
derivative terms from type IIB string theory are considered,
studying leading 't Hooft coupling corrections to ${\cal {N}}=4$ SYM
theory. Thus, although the Einstein-Hilbert action is supplemented
with the Weyl term (\ref{10d-corrected-action}), the entanglement
entropy is computed in terms of the same minimal surface used in the
RT prescription.

Now, let us consider Wilson loops. It is worth nothing that these
observables are not computed in terms of the supergravity action.
Instead, one directly uses the string theory action for curved
backgrounds \cite{Maldacena:1998im}. Then, $\alpha'$ corrections are
analyzed in a different way. They correct the background, but they
also demand to consider quadratic fluctuations around the classical
solution. Although the later give the leading corrections to the
Wilson loop (order ${\mathcal O}(\alpha')$ with respect to the
classical solution), they are seen as wrinkles of the classical
surface, and they are not expected to be relevant for the discussion
of thermalization times. A quantitative estimate of this claim is
out of the scope of the present work, nevertheless, our results
displayed on the figures in the following section strongly support
this statement. This is so, because the conclusions for
thermalization times obtained by using different observables are
nearly the same.

~

\textbf{Wightman two-point function}

~

Equal time two-point correlation functions of boundary quantum field
theory operators are related to space-like geodesics. In the AdS-BH
region of the bulk these are the ones that minimize the functional
length
\bea
{\mathcal L}=R \int_{-\frac{\ell}2}^{\frac{\ell}2} dx \frac{\sqrt{1-2
e^{\gamma(a(z)+5 b(z))} v'(x) z'(x)-e^{\gamma(2 a(z)+8 b(z))} f(z)
v'(x)^2}}{z(x)} \, ,
\eea
where $x$ is one of the $x^i$-coordinates on the boundary and the
rest have been chosen to be fixed along the curve. Since there is no
explicit dependence on the $x$ variable, the associated
``Hamiltonian'' is a constant of motion. It leads to the equation
\bea
1-2 e^{\gamma(a(z)+5 b(z))} \, v'(x) \, z'(x) - e^{\gamma(2 a(z)+8
b(z))} \, f(z) \, v'(x)^2 = \left(\frac{z^*}{z}\right)^2\label{CofM}
\, ,
\eea
where we have used the fact that $v(x)$ and $z(x)$ have a maximum at
$x=0$, and we have introduced $z^*=z(0)=z_{max}$.

On the other hand, the time independence of the metric in the AdS-BH
region requires $t'(x)=0$ as solution to the boundary conditions
$t(\frac{\ell}2)=t(-\frac{\ell}2)$. It leads to
\bea
v'(x)=-\frac{e^{-\gamma \left[a(z)+3b(z)\right]}}{f(z)}z'(x) \, ,
\eea
which together with equation (\ref{CofM}) can be used to express
$z'(x)$ as a function of $z$. Furthermore, one can use it to obtain
the following expression for the separation of the geodesic ends
\bea
\ell=2\int_{0}^{z^*} \frac{dz}{\sqrt{e^{-2\gamma
b(z)}\, f(z)\left[\left(\frac{z^*}{z}\right)^2-1 \right]}} \, .
\eea
Notice that this has been derived within the Einstein frame.

~

\textbf{Rectangular Wilson Loops}

~

Wilson loops are computed in terms of the Polyakov action for curved
spaces \cite{Maldacena:1998im}. Since we are working within the
Einstein frame, the dilaton appears in two different ways in the
action, namely: in a conformal factor connecting the string metric
with the Einstein metric, and also multiplying the world-sheet
scalar curvature. The later is order $\alpha'$ with respect to the
metric contribution, therefore, the leading $\phi_0$ term gives rise
to the string coupling constant while $\phi_1$ leads to a
sub-leading contribution with respect to the metric corrections, so
they are neglected here. Concerning the conformal factor, it is
actually a function of $\phi-\phi_0$, so it introduces only
$\phi_1(z)$ in the action.

Then, the minimal area surface related to space-like rectangular
Wilson loop with edges of length $\ell$ and $L$, satisfying
$L\gg\ell$, is determined by the functional
\bea
{\mathcal A}=R^2 L\int_{-\frac{\ell}2}^{\frac{\ell}2} dx \,
e^{\frac12\gamma \phi_1(z)}\frac{\sqrt{1-2 e^{\gamma(a(z)+5 b(z))}
v'(x) z'(x)-e^{\gamma(2 a(z)+8 b(z))} f(z) v'(x)^2}}{z(x)^2} \, .
\eea
Following similar steps as for the geodesic length case one finds
\bea
\ell=2\int_{0}^{z^*} \frac{dz}{\sqrt{e^{-2\gamma
b(z)}\, f(z)\left[\left(\frac{z^*}{z}\right)^4e^{\gamma(\phi_1(z)-\phi_1(z^*))}-1 \right]}} \, .
\eea

~

\textbf{Circular Wilson Loops}

~

The minimal area surface associated with a space-like circular
Wilson loop with radius ${\mathcal R}$ is given by
\bea
{\mathcal A}=2\pi R^2 \int_{0}^{\mathcal R} d\rho \frac\rho{z(\rho)^2}
e^{\frac12\gamma \phi_1(z)} \sqrt{1-2 e^{\gamma(a(z)+5 b(z))} v'(\rho) z'(\rho)
- e^{\gamma(2 a(z) + 8 b(z))} f(z) v'(\rho)^2} \, . \label{CWL}
\eea
Unfortunately, there is no constant of motion associated to $\rho$
in this case. Nevertheless, we can use the equation coming from the
condition $t'(\rho)=0$ to replace $v'(\rho)$ as a function of
$\left(z(\rho),z'(\rho)\right)$ in equation (\ref{CWL}), leading to the
following equation of motion
\bea
&&-\frac12 \gamma \partial_z \phi_1 \rho z(\rho) z'(\rho)\left(1-z(\rho)^{4}\right)
\left(1-z(\rho)^4+e^{2\gamma b} z'(\rho)^2\right)\cr
&&+\left(1-z(\rho)^4+e^{2\gamma b(z)}z'(\rho)\right)\left(2\rho\,
(1-z(\rho)^4) +e^{2\gamma b(z)} z(\rho) z'(\rho)\right)+ 2 \rho\,
e^{2\gamma b(z)} z(\rho)^4 z'(\rho)^2\cr &&+ \frac12
(1-z(\rho)^4) \rho\,
\partial_z\left(e^{2\gamma b(z)}\right) z(\rho) z'(\rho)^2
+(1-z(\rho)^4)\rho\, e^{2\gamma b(z)} z(\rho) z''(\rho)=0 \, .
\eea
No analytical solution is known for this non-linear differential
equation. Thus, expressions with ${\mathcal R}$ as a function of
$z^*$ are only numerically available.

~

\textbf{Entanglement entropy}

~

Now, we shall focus on the entanglement entropy of spherical regions
at the boundary theory. The minimal volume of the bulk hyper-surface
ending on it is specified by
\bea
{\mathcal V}=4\pi R^3 \int_{0}^{\mathcal R} d\rho \frac{\rho^2}{z(\rho)^3}
\sqrt{1-2 e^{\gamma(a(z)+5 b(z))} v'(\rho) z'(\rho) - e^{\gamma(2
a(z)+8 b(z))} f(z) v'(\rho)^2} \, .\label{EntE}
\eea
After following similar steps as for the circular Wilson loop case
one obtains the following differential equation
\bea
&&\left(1-z(\rho)^4+e^{2\gamma b(z)}z'(\rho)\right)\left(3\rho\,
(1-z(\rho)^4)+2e^{2\gamma b(z)} z(\rho) z'(\rho)\right) + 2 \rho\,
e^{2\gamma b(z)} z(\rho)^4 z'{}^2(\rho)\cr &&+ \frac12
(1-z(\rho)^4) \rho\,
\partial_z\left(e^{2\gamma b(z)}\right) z(\rho) z'{}^2(\rho)
+(1-z(\rho)^4)\rho\, e^{2\gamma b(z)} z(\rho) z''(\rho)=0 \, .
\eea
Again, ${\mathcal R}$ can be expressed as a function of $z^*$ only
numerically.

~

In the classical limit, where all these non-local observables are
well approximated as functions of these geometric objects, the
thermalization at a given scale is determined by the position of the
shell at that moment. If all non-local observables testing a
particular scale have their associated $z^*$ smaller than
$z(t)_{shell}$, we consider that this scale has thermalized. In the
next section we shall consider the thermalization time as a function
of the boundary scale, {\it e.g.} the separation between two
operator insertions at the boundary theory, between Wilson lines, or
entangled region, for the different non-local observables discussed
here once quantum corrections are taken into account.

%%%%%%%%%%%%%%%%%%%%%%%%%%%%%%%%%%%%%%%%%%%%%%%%%%%%%%%%%%%%%%
\section{Results and discussion}
%%%%%%%%%%%%%%%%%%%%%%%%%%%%%%%%%%%%%%%%%%%%%%%%%%%%%%%%%%%%%%

It is worth noting that the shell moves slower near the boundary as
$\lambda$ decreases from the strong coupling limit. However,
$z^*(\ell)$ and $z^*({\mathcal R})$ decrease in such a way that the
net result renders a faster thermalization. Also notice that the
opposite behavior is observed near the black hole horizon.

The results for the thermalization times are displayed in figure 1.
The sequence of curves: blue, green, brown and red correspond to
$\lambda\rightarrow\infty$, 150, 70, 30, respectively and $z_h=1$.
All non-local probes used to study thermalization show that there
exist a critical length, which at least for the range studied
($0.5<z_h<5$) is located about $(1.7-1.8)\times z_h$, for two-point
functions, circular Wilson loops and entanglement entropy, and about
$0.95\times z_h$ for rectangular Wilson loops. Shorter distances than
the critical scale measure a reduction in the thermalization time,
while larger distances perceive a delay. Figure 2 shows only the
$\alpha'^3$ corrections for case of the geodesics in order to
display more clearly the crossover point. Of course, as the UV modes
thermalize near instantaneously, their $\alpha'$ corrections are
hardly appreciable.  Another distinction between rectangular Wilson
loops and the rest of the observables is a delay in reaching the
equilibrium state (see figure 1.b). This effect is likely to be a
consequence of the fact that rectangular Wilson loops are not
completely contained in a region of size $\ell$, because of the
presence of the scale $L$, which is a second large scale for this
particular observable. This is what makes a difference with respect
to the other non-local observables discusses in this work. Notice
that although the geodesic lengths and minimal surfaces depend on
$R$, the thermalization times do not depend on $R$, so $z_h$ is the
unique relevant scale to this discussion.\footnote{It is worth
mentioning that the situation is different if one changes the
degrees of freedom of the shell. As discussed in \cite{Baron:2012fv}
and displayed in figure 2 of $loc.\,cit.$, when the shell is not
composed by a massless pressureless fluid, the velocity of the shell
depends on the radius of AdS space.}

\begin{figure}
\centering \subfigure[Two-point function]{
\includegraphics[scale=0.86]{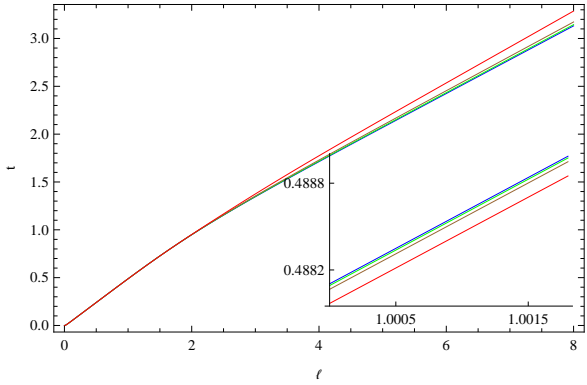}
\label{TwoPF} } \subfigure[Rectangular Wilson loop]{
\includegraphics[scale=0.85]{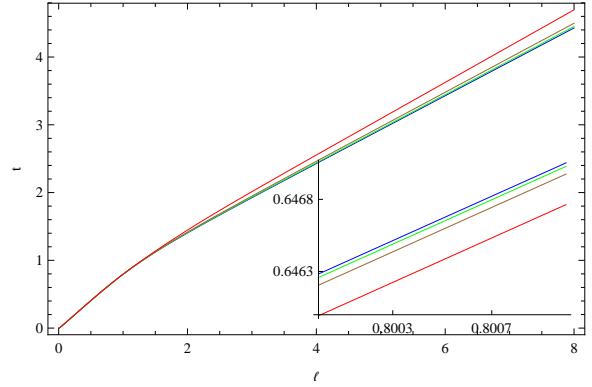}
\label{WilLR} }
\subfigure[Circular Wilson loop]{
\includegraphics[scale=0.86]{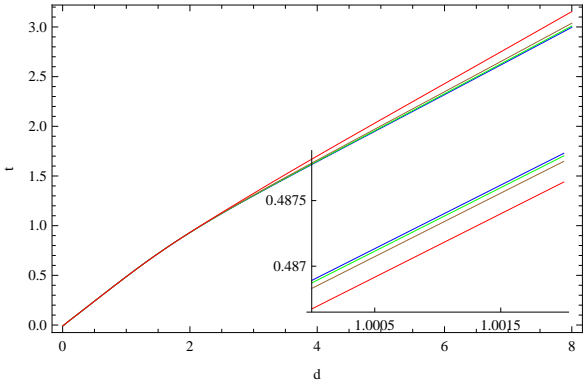}
\label{WilLC} } \subfigure[Entanglement entropy]{
\includegraphics[scale=0.86]{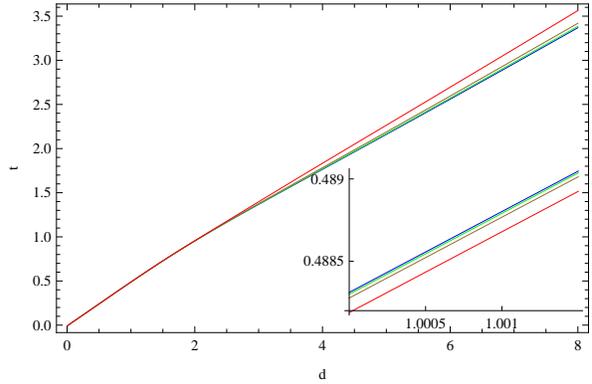}
\label{EntE} }
\caption{\footnotesize Figures show thermalization times as functions of
spatial resolution on the boundary theory. In figures (a) and (b),
$\ell$ denotes geodesic separation and separation for the
rectangular Wilson loop, respectively. In figures (c) and (d), $d=2{\cal R}$
denote the diameter of the circular Wilson loop and of the sphere
associated with the entanglement entropy. Blue curves describe the
thermalization pattern without $\alpha'$ corrections
($\lambda\rightarrow\infty$), the green ones correspond to
$\lambda=150$, brown to $\lambda=70$ and the red ones to
$\lambda=30$.} \label{tVsl}
\end{figure}

\begin{figure}
\centering \includegraphics[scale=0.65]{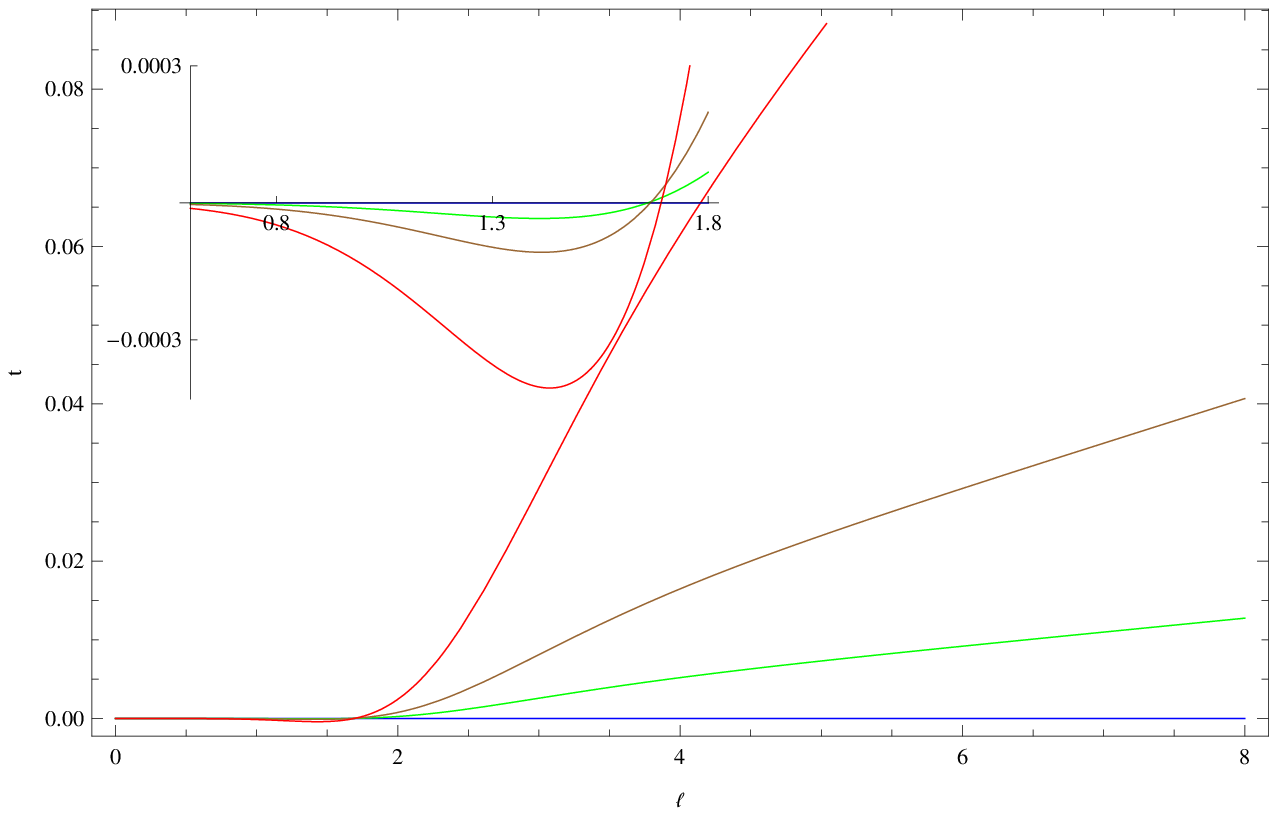}
\caption{\footnotesize This plot displays the $\alpha'$ corrections
to the thermalization time as a function of the scale resolution in
the case of two-point functions.  The identification among colors
and couplings is as in Figure 1.} \label{AlphCorr}
\end{figure}

We have considered corrections to thermalization time for systems
with the same amount of energy parameter $M$ at different strong couplings,
$\lambda$. Another interesting scenario is to compare thermalization
time for systems evolving to a given thermal equilibrium
temperature. By doing so, instead of a fixed $z_h$, it requires (see
Eq. (\ref{temp}))
\bea
z_h=z_h^{(0)}\rightarrow
z_h^{(0)}\left(1+\frac{265}{16}\gamma\right) \, .
\eea
In this situation, the thermalization is still very similar to the
previous case, the only difference is that the crossover point is
shifted to the IR and is located at about $(1.9-2.1)\times z_h$.

Although we are analyzing an ${\cal {N}}=4$ SYM plasma, one is
tempted to compare the thermalization patterns with more realistic
scenarios like strongly coupled QGP of RHIC and LHC experiments. In
these situations two heavy ions collide at relativistic velocities
and after that thousand of particles are created and a fraction of
the kinetic energy of the original nuclei is converted into heat
leading to a finite temperature strongly coupled plasma.

It means that in order to analyze more rigorously the pattern of
thermalization times with large yet finite coupling, one should be
able to model the amount of kinetic energy which is converted into
heat during the collision, when the coupling decreases from infinity
to a given finite value. This information is an input for our
approach which enters the mass of the collapsing shell. The
knowledge of this information in the case of QGP requires that one
should be able to model the whole process: collision, deconfinement
and thermalization. However, since ${\cal {N}}=4$ SYM is not a
confining quantum field theory, it is not expected that this
information could be inferred from our analysis.

Numerical analysis shows that if the rate of kinetic energy
transferred increases, the critical scale shifts towards the IR (as
for instance in the fixed-temperature scenario). On the other hand,
if it decreases ($z_h$ increases) there exist a bound
$z_h\rightarrow z_h(1+\Lambda\gamma),~\Lambda\lesssim 20$, where
there is still a critical length, and above this limit any scale
experiences a delay in reaching the thermal equilibrium.

We have not displayed here the results with $1/N$ corrections.
Nevertheless, it is easy to see that there is no qualitative
difference. In fact, there is only a hardly appreciable difference
with respect to the case with $\alpha'$ corrections. The effective
value for $\gamma$, when both $\alpha'$ and $1/N$ corrections are
considered, is given by (see equation (\ref{ModForm}))
\bea
\gamma=\frac18 \zeta(3)\frac1{\lambda^{\frac32}}~\rightarrow
~\frac18\, \frac1{\lambda^{\frac32}}\left(\zeta(3) +
\frac1{48}\left(\frac\lambda N\right)^2 + \frac{1}{2 \pi^{1/2}}
\left(\frac\lambda N\right)^{\frac32} e^{-8\pi^2 \frac
N\lambda}\right) \, .
\eea
Therefore, both $\alpha'$ and $1/N$ corrections go in the same
direction. Both types of corrections increase the value of $\gamma$,
however, since we are working in the limit where $\lambda\ll N$, the
$1/N$ contributions can be seen as sub-leading ones \footnote{This
holds for geodesic lengths. On the other hand, for Wilson loops the
leading $1/N$ corrections due to world-sheet with handles have been
discussed by Drukker and Gross in \cite{Drukker:2000rr}, and they
turn to be also additive. Also notice that $\alpha'$ corrections to
circular Wilson loops due to string fluctuations were discussed by
Forste, Ghoshal and Theisen in \cite{Forste:1999qn} at zero
temperature.}.

%%%%%%%%%%%%%%%%%%%%%%%%%%%%%%%%%%%%%%%%%%%%%%%%%%%%%%%%%%%%%%
\section{Conclusions}
%%%%%%%%%%%%%%%%%%%%%%%%%%%%%%%%%%%%%%%%%%%%%%%%%%%%%%%%%%%%%%

We have studied type IIB stringy theory corrections to
thermalization processes in strongly coupled SYM plasmas. Our
results show that both $\alpha'$ and $1/N$ contributions lead to
delays in thermalization times for IR modes, while they make
slightly faster thermalization time for UV modes, both in comparison
with the strong coupling limit, when comparing equal temperature or
equal energy scenarios. This statement is based on the study of the
time evolution of extended geometric probes in the bulk which are
connected to specific non-local observables in the boundary field
theory plasma.

We find that finite 't Hooft coupling corrections decrease very
little the thermalization time of UV modes, while they produce the
opposite trend for IR modes, which thermalize slightly later in
comparison with their corresponding behavior at the strong coupling
limit. Thus, the observed trend is that the leading string theory
corrections enhance the difference of thermalization time between UV
and IR modes.

In the figures shown we consider the situation where the black hole
horizon does not change by the effects of string theory corrections.
Equation (\ref{temp}) indicates that the equilibrium temperature increases
as the 't Hooft coupling decreases from the strong coupling limit. Besides,
we also have numerically investigated the case when the equilibrium
temperature is kept fixed, and then the black hole radius becomes
smaller. In this case our results agree with the pattern indicated
in the figures.

Although this approach has no access to the information about
coupling corrections to the rate of kinetic energy converted into
thermal energy during the heavy-ion collision, it is indeed
interesting to study which are the consequences of varying such a
conversion rate. We numerically show that if this rate increases,
the critical length shifts toward the IR. On the other hand, if it
decreases, there is a bound above which there is no crossover and
the obtained pattern shows a delay at any scale.

Besides, it has been recently discussed the effect of certain
aspects of the thermalization time scale in terms of finite values
of the 't Hooft coupling, but from a very different perspective. In
particular, in references \cite{Steineder:2012si,Steineder:2013ana}
it has been investigated the effect of finite values of the 't Hooft
coupling in comparison with the $\lambda \rightarrow \infty$ limit,
using a particular version of the holographic thermalization within
a quasi-static approach, by looking at spectral densities to infer
the thermalization times for UV and IR modes. Recall that spectral
densities are obtained from two-point correlation functions of
electric currents in the plasma.\footnote{Particularly, for the
$SU(N)$ ${\cal {N}}=4$ SYM plasma at thermal equilibrium the
current-current correlators, spectral densities, photoemission and
lepton pair production rates were firstly obtained in
\cite{CaronHuot:2006te}. In order to include the full ${\cal
{O}}(\alpha'^3)$ corrections one has to consider not only the effect
of the term quartic in the ten-dimensional Weyl tensor, but also the
corrections coming from the Ramond-Ramond 5-form field strength
$F_5$. These effects can be collected into the ${\cal {T}}$-tensor
mentioned above. This leads to a very complicated series of
calculations which have been carried out in \cite{Hassanain:2009xw1}
within the deep inelastic scattering regime, in
\cite{Hassanain:2009xw2} within the hydrodynamic regime, in
\cite{Hassanain:2010fv} for the electrical conductivity of the
plasma, in \cite{Hassanain:2011ce,Hassanain:2012uj} for the
photoemission rates in the thermally equilibrated SYM plasma.}
Whenever current correlators are studied in the $SU(N)$ ${\cal
{N}}=4$ SYM plasma, it is necessary to introduce an external gauge
field in the boundary field theory to be coupled to the current. The
currents are to be coupled to a vector fluctuation of the bulk
metric, thus turning on a contribution from the Ramond-Ramond
five-form field strength in the ${\cal {O}}(\alpha'^3)$ corrections
to the type IIB string theory action. These contributions produce
very large corrections in SYM plasma observables associated with
electric charge transport, such as electrical conductivity, and
photo-emission rates. This could be one of the reasons for the
difference with respect to our work where no external Abelian gauge
field is considered. Another difference is that if one uses
non-local observables associated with charged particles in the
boundary field theory, in order to probe thermalization at a given
scale, not only the position of the shell is relevant but also the
photon spectrum from the plasma has to be taken into account. In
addition, another important difference is that in the approach of
\cite{Steineder:2012si,Steineder:2013ana} the idea is to consider a
quasi-static shell rather than a dynamical one, and look at certain
positions of this shell, which are close to the event horizon of the
black hole. From the point of view of the thermalizing plasma this
situation corresponds to consider a SYM plasma near the thermal
equilibrium. This situation is very different compared with a
dynamical holographic thermalization process because in the
dynamical case one follows the collapse of the thin shell starting
from the boundary of the AdS space, {\it i.e.} completely far away
from the black hole horizon, and this corresponds to a SYM plasma
which is very far from its thermally equilibrated state. This last
case seems to be somehow closer to a realistic plasma thermalization
process initiated by a heavy ion collision than the quasi-static
approach.

It is worth noting that some of the conclusions displayed here could
be altered if one considers two-point functions of operators with no
large conformal weights. Another crucial ingredient in our work was
the assumption of Vaidya as a solution of type IIB supergravity on
$AdS_5\times S^5$ background. Nevertheless, as commented in
\cite{Bhattacharyya:2009} the pressureless null dust is only
understood as an approximation in the field configuration space. For
this reason it would be very interesting to consider quantum
corrections in the case of more general shells, for instance those
discussed in \cite{Baron:2012fv} by generalizing the Israel junction
conditions in higher derivative gravity theories. By doing so, one
should be able to see if the pattern on thermalization times
observed here is generic, or it is just a consequence of the degrees
of freedom chosen for the shell.

~

%%%%%%%%%%%%%%%%%%%%%%%%%%%%%%%%%%%%%%%%%%
\centerline{\large{\bf Acknowledgments}}
%%%%%%%%%%%%%%%%%%%%%%%%%%%%%%%%%%%%%%%%%%

~

We thank Dami\'an Galante for early collaboration on the subject and
for a critical reading of the manuscript. We also thank Robert Myers
for valuable discussions concerning entanglement entropy in higher
derivative gravity, and to Carlos N\'u\~nez and Guillermo Silva for
interesting discussions about Wilson loops. This work has been
supported by CONICET, the Consejo Nacional de Investigaciones
Cient\'{\i}ficas y T\'ecnicas of Argentina, and the ANPCyT-FONCyT
Grant PICT-2007-00849. The work of M.S. has also been supported by
the CONICET Grant PIP-2010-0396.

\end{document}